\documentclass[12pt]{article}
\usepackage{graphicx}
\usepackage{picinpar}
\newcommand{\ba}{\begin{eqnarray}}
\newcommand{\ea}{\end{eqnarray}}
\newcommand{\be}{\begin{equation}}
\newcommand{\ee}{\end{equation}}

\begin{document}
\begin{titlepage}
\begin{flushright}
hep-ph/0108032{\hskip.5cm}\\ IOA-TH.10/01\\ NTUA 06/01
\end{flushright}
\begin{centering}
\vspace{.3in}
{\bf On the stability of nontopological solitons in supersymmetric}\\
{\bf  extensions of the Standard Model: A Numerical Approach}\\
\vspace{2 cm}
{G.K. Leontaris$^1$, A. Prikas$^2$, A. Spanou$^2$,
N.D. Tracas$^2$,
 N.D. Vlachos$^3$} \\
\vspace{1 cm}
{\it {
$^1$Physics Department,
University of Ioannina, Ioannina 45 110, Greece}}\\
{\it {
$^2$Physics Department, National Technical University,
       Athens 157 73 , Greece}}\\
{\it {
$^3$Physics Department, University of Thessaloniki,
       Thessaloniki 540 06, Greece}}\\
\vspace{1.5cm}
{\bf Abstract}\\
\end{centering}
\vspace{.1in}

\noindent
In various supersymmetric extensions of the Standard Model there
appear  non-topological solitons due to the existence
of $U(1)$ global symmetries associated with  Baryon and/or
Lepton quantum numbers.  Trilinear couplings (A-terms) in the scalar
potential break explicitly the $U(1)$ invariance. We investigate
numerically the stability  of these objects in the case that this
breaking is small. We find that stable configurations, oscillating
though, can still appear. Other relevant properties are also examined.
\vfill \hrule width 6.7cm
\begin{flushleft}
July 2001
\end{flushleft}

 \end{titlepage}

Non-topological solitons are non-dissipative solutions of finite
energy which arise in field theories possessing continuous  global
symmetries~\cite{Friedberg:1976me},\cite{Lee:1992ax},\cite{Coleman:1985ki}.
Of particular interest are possible stable configurations which
carry baryonic or leptonic charge~\cite{Kusenko:1997ad} and appear
in supersymmetric extensions of the Standard Model. In this latter
context, they appear to have interesting cosmological consequences
since they could be related to the baryon number of the universe,
or even they could play a role as dark matter
candidates~\cite{Kusenko:1998si}. In supersymmetric theories, the
trilinear superpotential couplings of the matter and Higgs
superfields, as well as the supersymmetry breaking terms, generate
a scalar potential which may possess exact or approximate global
symmetries which could be correlated to the above quantum numbers.
It is the purpose of this letter to examine a generic form of such
a scalar potential and investigate the properties of the solitonic
configurations.

In general, the resulting scalar potential in a supersymmetric
theory depends on various scalar fields, $U=
U(\Phi_1,\Phi_2,\dots, \Phi_n)$, however, in the present work we
will concentrate on the simplest case of only one field $\Phi$,
i.e. $U= U(\Phi)$, while the generalization is straightforward. The
contributions from supersymmetry breaking and non-renormalizable
terms in the scalar potential have the generic form~\footnote{see for example
~\cite{Enqvist:1999en},\cite{Axenides:1999dw}.}
\ba
U(\Phi)&=&m_s^2|\Phi|^2+\sum_n\left(\lambda_{NR}^2\frac{|\Phi|^{2
(n-1)}}{M^{2(n-3)}}+ A
\frac{\Phi^n}{M^{n-3}}+A^*\frac{\Phi^{*n}}{M^{n-3}}\right)
\label{pote}
\ea
In (\ref{pote}), $m_s$ is a soft mass term, of the order of
the supersymmetry breaking  scale. $\lambda_{NR}$ is a dimensionless
constant  associated to the Yukawa coupling of the corresponding
non-renormalizable term of the superpotential. $A$ is the coefficient
of the trilinear term in the scalar potential (the so-called $A$-term),
while $M$ is of the order of the unification scale.
It is easily observed~\cite{Axenides:1999dw} that in the case of a
scalar potential  with vanishing $A$-terms the
global $U(1)$ symmetry of the Lagragian survives. Then, for
appropriate values of the soft supersymmetry breaking parameters
and Yukawas  which determine the coefficients of the various
potential terms, a stable solitonic configuration does appear. However,
the $A$-terms are usually non-vanishing. Even if their initial
value is zero, renormalization group running will generate a
non-vanishing value and the aforementioned global $U(1)$
symmetry breaks. In this letter we wish to investigate the
stability of the solitonic configuration in the presence of small
non-zero $A$-terms which induce  small perturbative terms in the
scalar potential.

 To this end, we will use a simplified model in $1+1$ dimensions
and adopt the conventions and notations of
reference~\cite{Axenides:2000hs}. Assuming a complex scalar filed
$\Phi$, the dynamics of the resulting scalar field theory can be
described by a Lagrangian density of the form
\be
\label{lagrangian} {\cal L}= \frac 12
\partial_{\mu}\Phi\partial^{\mu}\Phi^*-U(\Phi)
\ee Starting  with a potential respecting a $U(1)$ invariance
\begin{equation}\label{potential}
U(\Phi)=\frac12 m^2|\Phi|^2-\frac13\alpha|\Phi|^3+\frac14
b|\Phi|^4
\end{equation}
while making use of the rescalings~\cite{Axenides:2000hs}
\begin{equation}\label{rescaling}
\Phi\rightarrow\frac{m^2}{\alpha}\Phi,\quad\quad
x\rightarrow\frac{x}{m}
\end{equation}
one obtains  the Lagrangian
\begin{equation}
\label{r-lagrangian}
\mathcal{L}=\frac12  \partial_\mu \Phi^*\partial^\mu \Phi-
\frac12|\Phi|^2+\frac13|\Phi|^3-\frac14 B|\Phi|^4
\end{equation}
with the parameter $B =\frac{bm^2}{\alpha^2}$. Note that in order
to get the above Lagrangian we have divided by a factor $m^6/a^2$
which has dimensions of $m^2$. Therefore, the Lagrangian in
(\ref{r-lagrangian}) is
dimensionless. The equation of motion reads
\begin{equation}\label{eqmotion}
\ddot{\Phi}-\Phi''+\Phi-|\Phi|\Phi+B|\Phi|^2\Phi=0
\end{equation}
Using the ansatz~\cite{Coleman:1985ki}
\begin{equation}\label{qball}
\Phi(x)=\sigma(x)e^{i\omega t}
\end{equation}
and inserting  (\ref{qball}) in (\ref{eqmotion}) the following
differential equation for $\sigma(x)$ is obtained
\begin{equation}\label{seq}
\sigma''+(\omega^2-1)\sigma+\sigma^2-B\sigma^3=0
\end{equation}
The requirement for a finite energy configuration and the
asymptotic behaviour at infinite ``time'', $x\rightarrow \infty$,
imply the following conditions for $\sigma$:
\begin{equation}\label{bc}
\sigma'(0)=0, \quad\quad \sigma(\infty)=0.
\end{equation}
 The constraints imposed on $\omega$ by these conditions
are written as
 \be
 \label{constraint} 0<1-\frac{2}{9B}<\omega^2<1
\ee
The conserved charge is
\begin{equation}\label{charge}
Q=\frac{1}{2i}\int(\Phi^*\partial_0\Phi-\Phi\partial_0\Phi^*)dx
\end{equation}
Given the ansatz (\ref{qball}) one finds that the conserved charge
is given by
\begin{equation}
\label{qcharge}
Q= \omega\int \sigma(x)^2 d\,x
\end{equation}
{}Finally, the energy density is
\be
\label{energy} \mathcal{E}=\frac 12|\dot{\Phi}|^2+\frac
12|\Phi'|^2 +\frac 12|\Phi|^2-\frac 13|\Phi|^3+\frac 14 B|\Phi|^4
\ee
which, with the use of (\ref{qball}), becomes
\be
\label{qenergy}
\mathcal{E}=\frac 12\sigma'^2+\frac
12(1+\omega^2)\sigma^2-\frac13\sigma^3+ \frac 14 B\sigma^4
\ee
 A~~numerical analysis of the model shows that static field
configurations   are obtained for a large portion of the parameter
space $\omega^2, B$. Finally, the extension in $2+1$
dimensions shows that
similar static stable `objects' do exist, at least for certain
$\omega^2$-values~\cite{Axenides:2000hs}.

We wish now to add a $U(1)$ breaking term in the (rescaled) Lagrangian
(\ref{r-lagrangian}) of the form
\begin{equation}
\label{breaking_t}
\frac{s}{n!}(\Phi^n+\Phi^{*n})
\end{equation}
and examine in a similar manner the stability of the above
configurations. We  choose to work with $n=2$ and find that the
equations of motion for the scalar field in this case generalize
as follows
\ba
\ddot\Phi-\Phi''+\Phi-|\Phi|\Phi+B|\Phi|^2+2s\Phi^*&=&0
\label{mem1}
\\
\ddot\Phi^*-{\Phi^*}''+\Phi^*-|\Phi|\Phi^*+B|\Phi|^2+2s\Phi&=&0
\label{mem2}
\ea
The current is given by
$\Phi^*\partial_{\mu}\Phi-\Phi\partial_{\mu}\Phi^*$  and the
charge
$Q\propto\int(\Phi^*\partial_0\Phi-\Phi\partial_0\Phi^*)d\,x$,
respectively, however, due to the extra $U(1)$-breaking term the
charge is no longer conserved. Its rate of change is given by
\ba
\frac{d\,Q}{d\,t}&=&\int i s
(\Phi^2-\Phi^{*2})\,d\,x
\ea
An analytic solution of the above system is hard to find, so  our
intention is to solve numerically the complete equations of
motion, taking as starting point the `unperturbed' solution (\ref{qball}).

We first check that  our numerical methods reproduce the original
results in the case of vanishing $s$-term ( the $U(1)$-invariant
Lagrangian). Indeed, we have checked that the soliton-profile (the
absolute value of the solution),  to a very good
approximation,  remains constant in time while the charge,
energy and energy/charge
are constant within 2\%, 1.4\% and 0.02\% correspondingly.

\begin{figure}[t]
\centering
\begin{tabular}{ccc}
\includegraphics[scale=0.6]{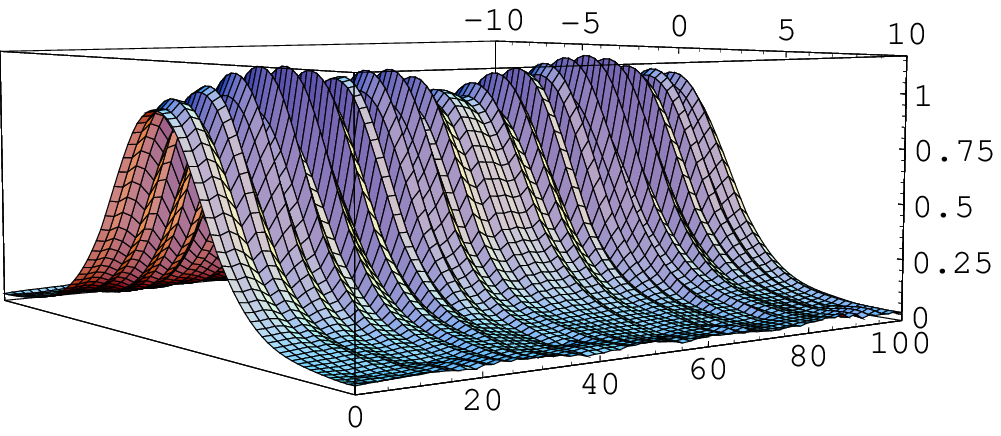}&
\includegraphics[scale=0.4]{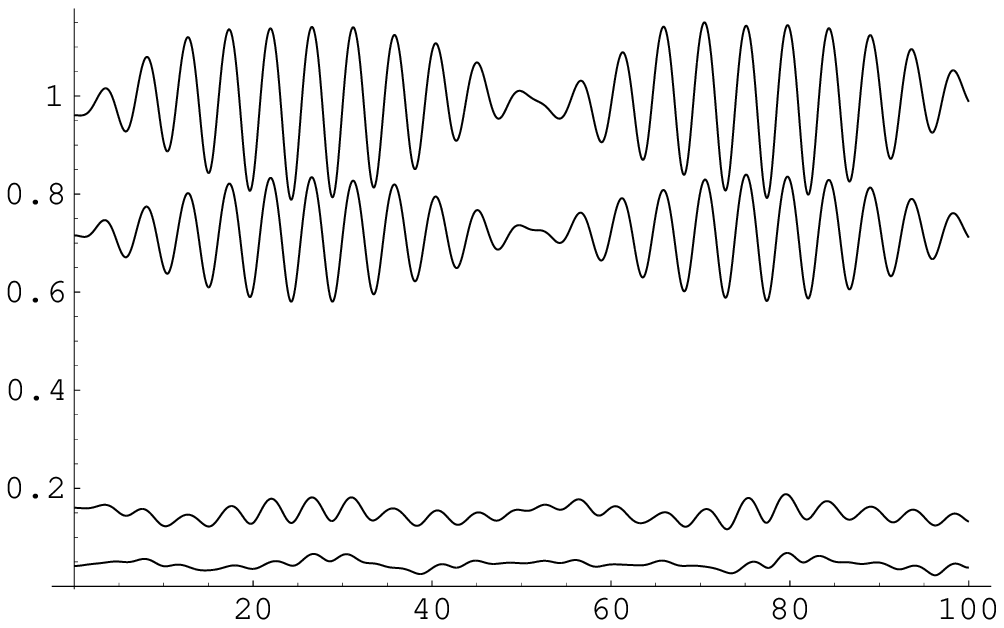}&
\includegraphics[scale=0.3]{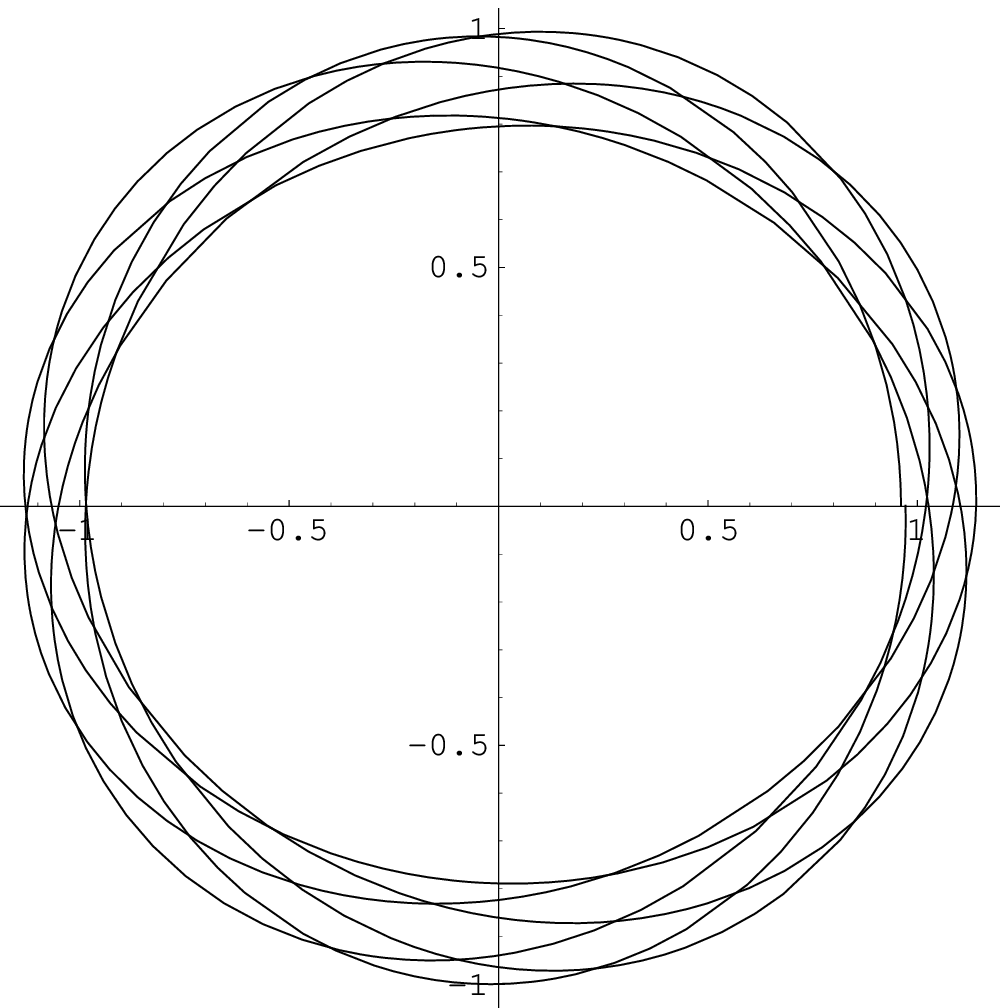}\\
\includegraphics[scale=0.6]{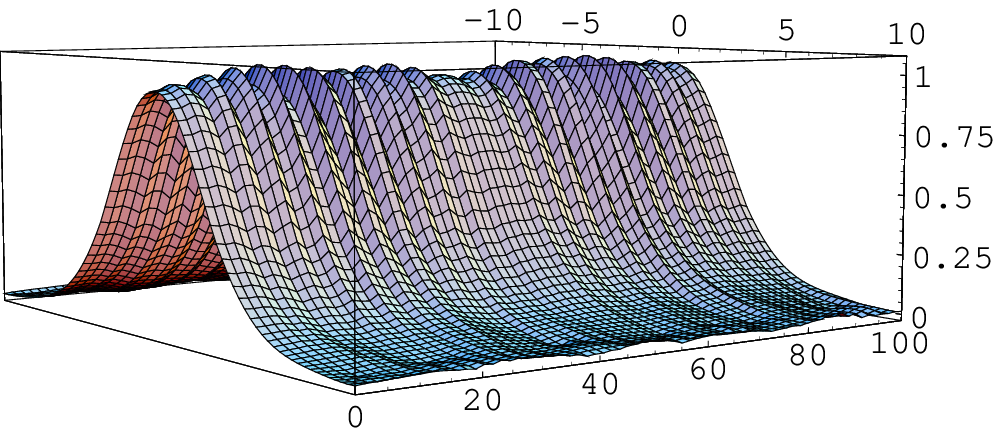}&
\includegraphics[scale=0.4]{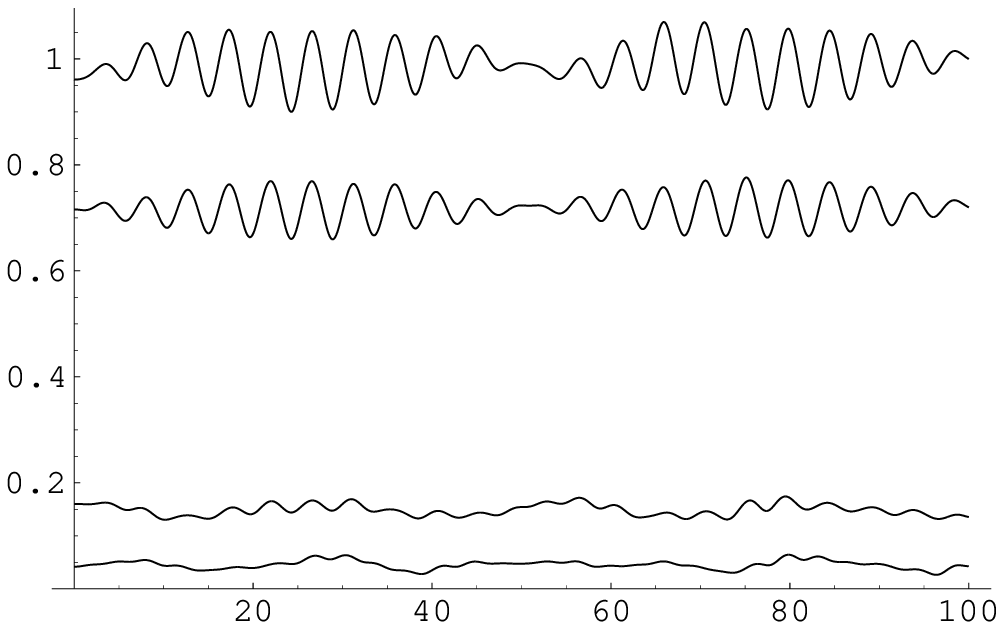}&
\includegraphics[scale=0.3]{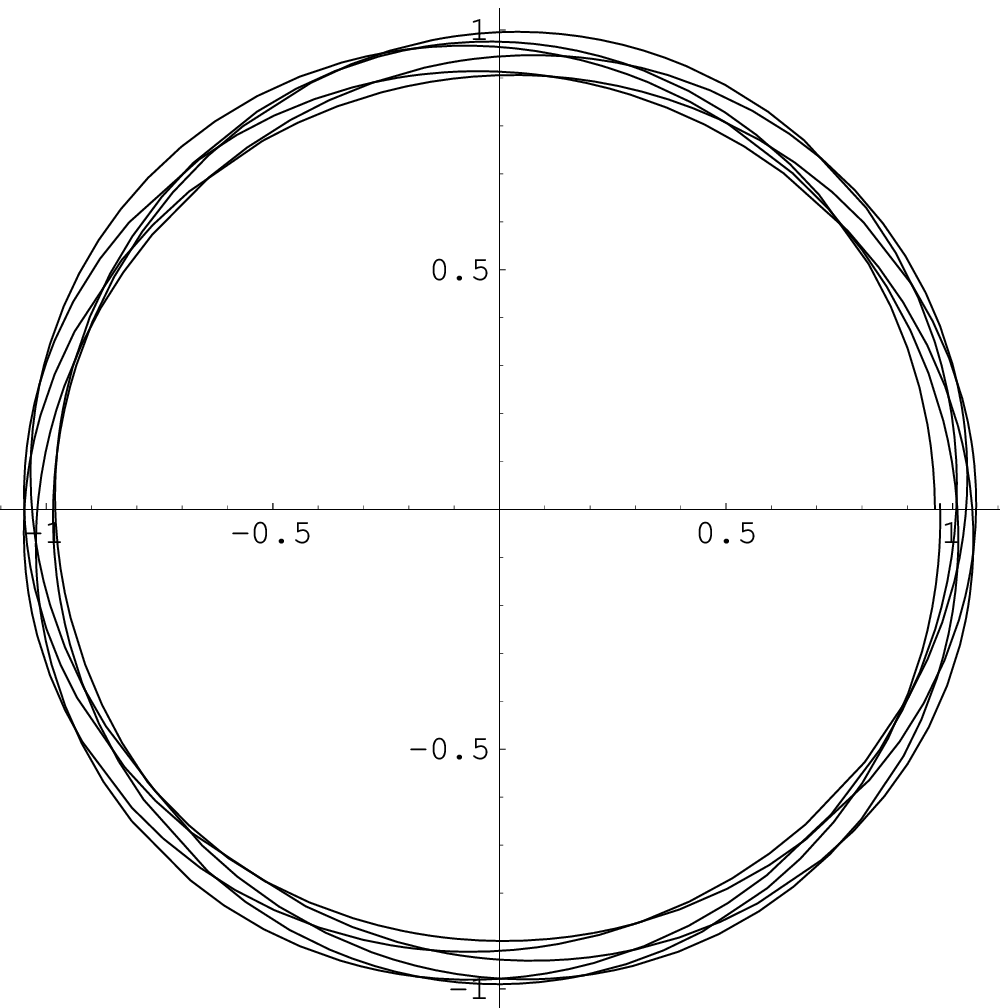}\\
\includegraphics[scale=0.6]{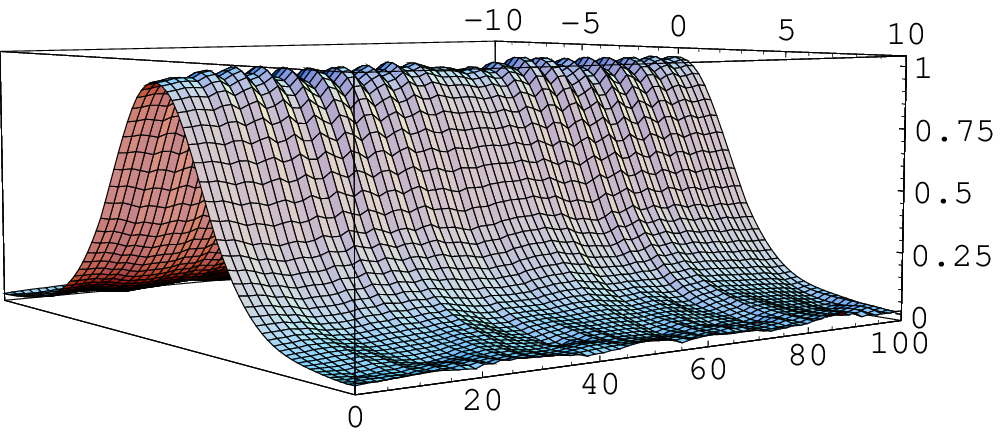}&
\includegraphics[scale=0.4]{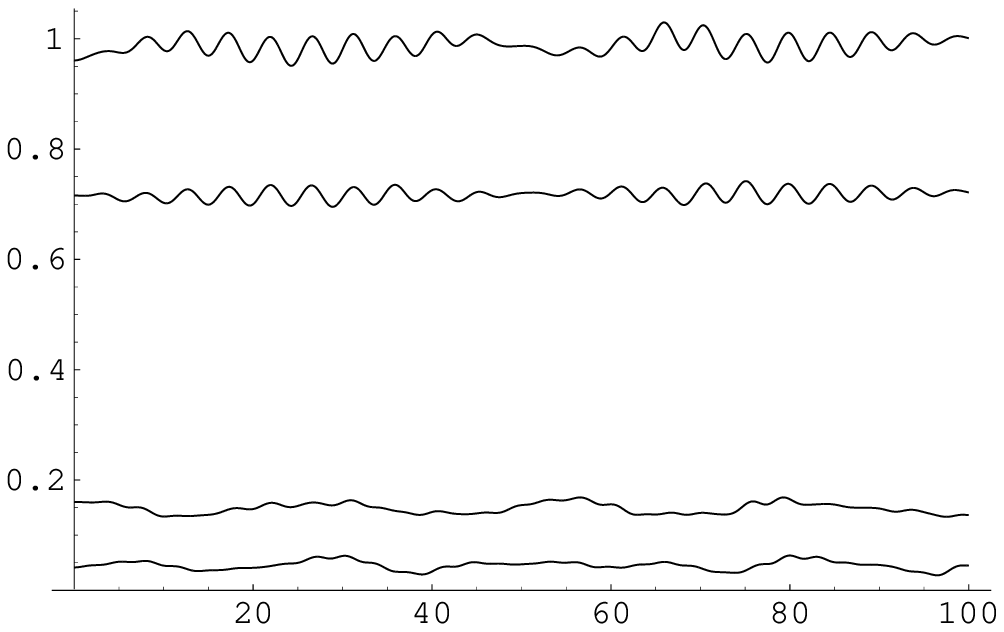}&
\includegraphics[scale=0.3]{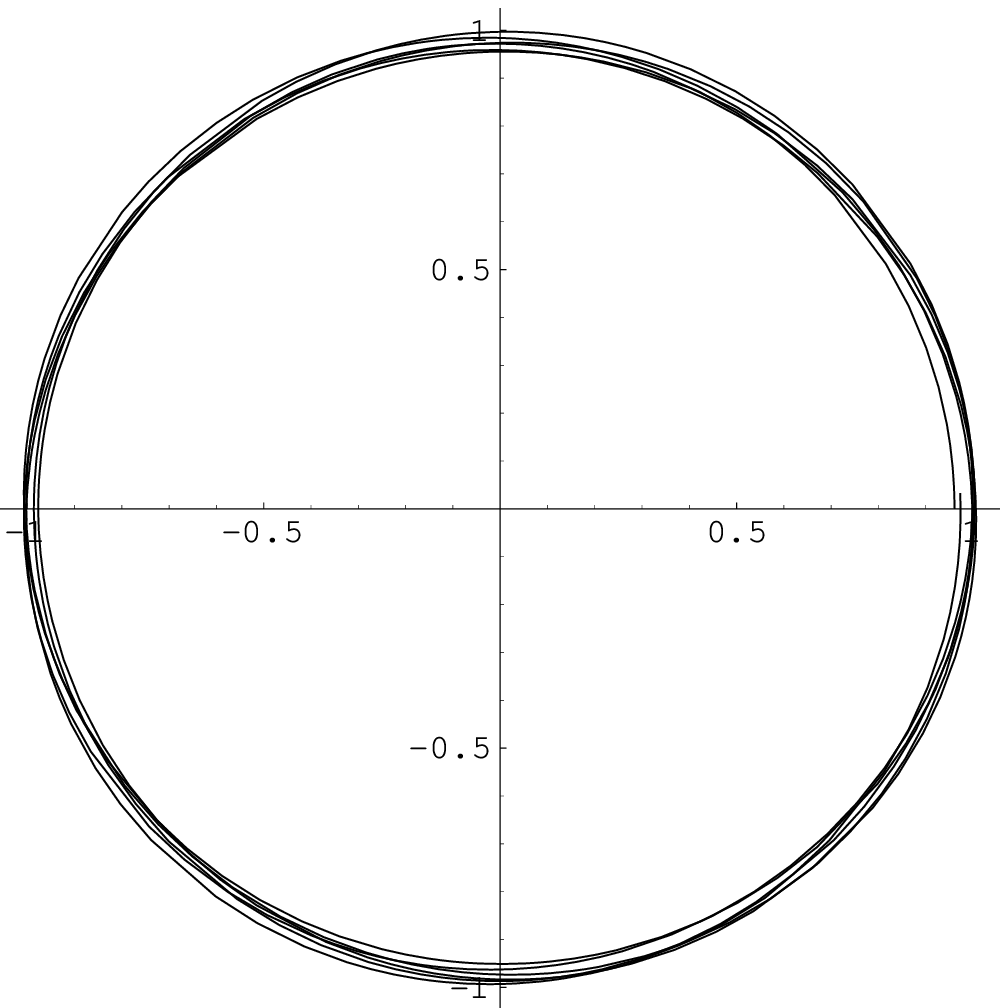}
\end{tabular}
\caption{The time evolution of the soliton profile, the
oscillation of the points $x=0,2,5,7$ and the parametric plot
(Real Part, Imaginary Part) for one period and for three values of
the breaking parameter $s=0.007,0.003,0.001$.} \label{fig1}
\end{figure}

Next, we switch on the symmetry breaking term in the potential. In
our numerical investigations   we use the values $B=0.34$ and
$\omega^2 =0.51$ for the two potential parameters. In
Fig.(\ref{fig1}) we show  the evolution of the profile in time,
the oscillation of $|\Phi(x,t)|$ in time for four points
($x=0,2,5,7$) ($x=0$ is the top of the profile), and finally the
parametric plot (Real Part, Imaginary Part) for the top of the
profile. The last one is drawn for one period of the beat-like
graph shown in the middle figure. The three rows correspond to
three values of $s=0.007,0.003,0.001$.

Some comments are in order. Although the $U(1)$ is explicitly
broken, we see that we have a stable configuration though
oscillating in time. We have checked the stability for a time
corresponding to at least 10 periods ($2\pi/\omega$). All points of the
profile oscillate in phase, while the ``movement'' of a point
resembles a beat. The higher we are on the solitonic profile the
larger the oscillation is. The parametric plot, which for a
soliton is a circle, transforms to a ``turning ellipse'' which
closes as a beat period is completed. As $s$ gets smaller, the
oscillation decreases, the top of the profile becomes smooth with the
time while in the parametric plot, the ``turning ellipse'' becomes
a circle. The period of the beat remains constant (within numerical
errors). The non-smooth ending of the profile for large $|x|$  is
also due to numerical errors in the solution of the equation of
motion.

\begin{figure}[t]
\centering
\begin{tabular}{cc}
\includegraphics[scale=0.6]{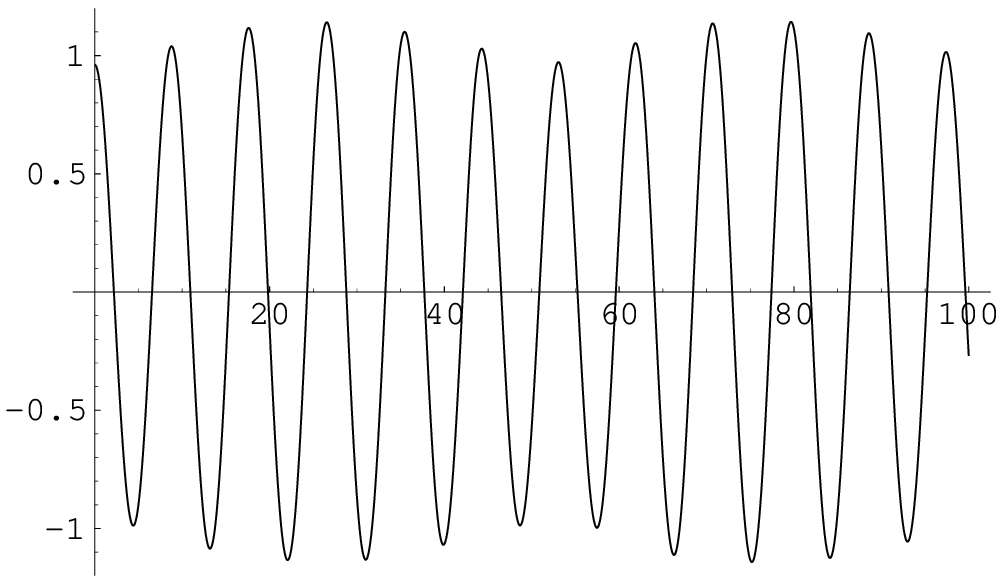}&
\includegraphics[scale=0.6]{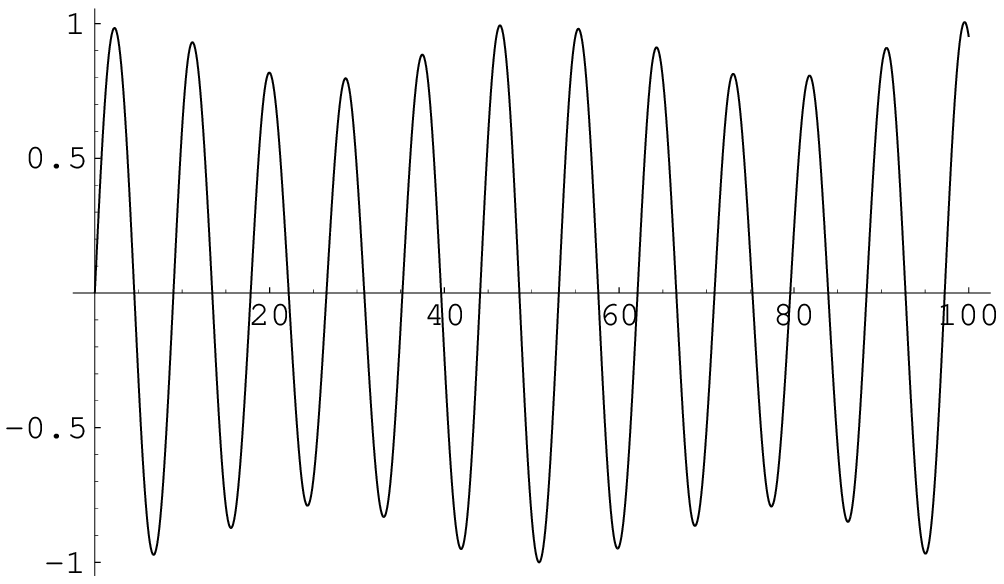}
\end{tabular}
\caption{The real (left) and imaginary part of the solution
as a function of time, for $x=0$ and $s=0.007$.}
\label{fig2}
\end{figure}

It is interesting to analyze the real and the imaginary part of
the solution, for $x=0$, as a function of time; this is shown in
Fig.(\ref{fig2}).
We see a clear oscillation with frequency $\omega$ which is
modulated by a lower frequency. We clearly see the cosine and the
sine function corresponding to the real and imaginary part
(one should bear in mind that the  ansatz for the solitonic configuration is
$\sigma(x)\exp(i\omega t)$). The modulating signal has  half a
period phase difference between real and imaginary part. It is
easy now to see that the high frequency of the beat-like profile
is $2\omega$.

As the breaking parameter $s$ gets bigger, the oscillations show
large amplitudes and the sense of a period fails to appear. This
loss of clear periodicity starts to appear when the ratio
energy/charge, which also oscillates with time, becomes
larger than one, which is the constraint for a stable
configuration against decaying to the fundamental mesons of the
theory.

\begin{figure}[t]
\centering
\includegraphics[scale=0.6]{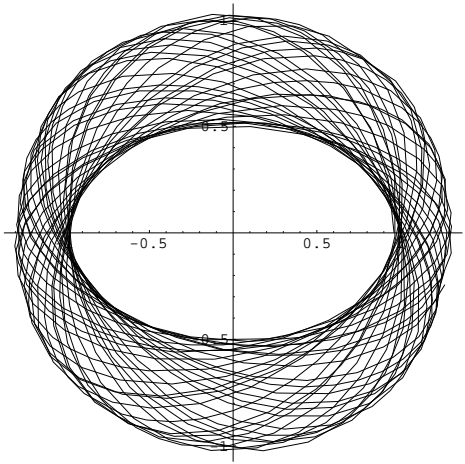}
\caption{Parametric plot (Real Part, Imaginary Part),
for $x=0$, $s=0.015$ and for $t=300$}
\label{fig3}
\end{figure}

At this point, we should explain what we mean by charge. In the
$U(1)$ invariant theory the conserved charge is defined by the
Eq.(\ref{charge}). We continue to define the charge by the same
equation, in the sense that our symmetry breaking parameter is
small. The energy is given by  Eq.(\ref{charge}), where of course
we have added the term coming from the breaking, namely $(1/6)s
Re[\Phi^2]$. The lack of periodicity is most clearly seen in the
parametric plot (Real Part, Imaginary Part) where now the line
tends to cover all the plane, exhibiting two ``fixed points''. In
Fig(\ref{fig3}) we show the parametric plot for $x=0$, $s=0.015$
and for $t=300$.
The ``fixed points'' could be attributed to a left over $Z_2$ symmetry.
Indeed, the symmetry breaking term
$\Phi^n+\Phi^{*n}$ shows a $Z_n$ symmetry if $\Phi=\sigma(x)\exp(i\omega t)$.

We have made a Fourier analysis of the real and imaginary part of
the complete solution for $x=0$ and, as it is obvious from the Figures, we
get large amplitudes just for two frequencies which are multiplets
of the basic frequency ($2\pi/T$, where $T$ is the period of the
beat) namely $\omega$, which has the largest amplitude and
$\omega'=\omega\pm(2\pi/T)$. To a very good approximation we can
write
\begin{equation}\label{solution}
\Phi(x=0,t)=c_R\cos(\omega t)+c'_R\cos(\omega' t)+
           i(c_I\sin(\omega t)+c'_I\sin(\omega' t))
\end{equation}
where the $c$'s are constant coefficients.
Since all points oscillate in phase we can write for the complete
solution
\begin{equation}\label{csolution}
\Phi(x,t)=K(x)\Phi(0,t)
\end{equation}
where $K(x)$ is an unknown function of the position. The first
obvious choice for $K(x)$ is $\sigma(x)$ itself. We have tried
therefore the solution
\begin{equation}\label{ccsolution}
\Phi(x,t)=\frac{\sigma(x)}{\sigma(0)}\Phi(0,t)
\end{equation}
which for the central region of the  profile differs from the
numerical solution by as low as 2\%. Fourier analysis can also
help to understand the transition to the simple  solution,
Eq.(\ref{qball}). As $s\rightarrow 0$, the amplitude of the
$\omega'$ frequency gets smaller and the only surviving frequency
is the $\omega$.

The above analysis holds when the parameters $B$ and $\omega$ of
the solution are deep inside the stability region of the
$(B,\omega^2)$ space. The stability region is defined from the
Eq.(\ref{constraint}) \cite{Axenides:2000hs}. When we are near the
boundary, the situation gets extremely complicated. Even with
a relatively small breaking parameter $s$, the energy/charge ratio
 becomes larger
than 1. The period of the beat is no more constant but depends on
the value of $s$. The Fourier analysis of the solution shows that
more than two frequencies have large amplitudes and, for large
enough $s$, none of them is equal to $\omega$. As $s\rightarrow
0$, all the amplitudes, except one, tend to zero, while the
remaining one corresponds to a frequency which tends to $\omega$
(since the beat period is not constant, the basic frequency of the
Fourier analysis, and therefore its multiplets also change).

We get a more clear situation near the boundary of the stable solutions,
in the case of the so called thin wall approximation: the function
$\sigma(x)$ has a constant value for a certain region in $x$ and
zero elsewhere. In that case, the Euler-Lagrange
equation, the current and the energy density of the soliton are given
respectively:
\begin{equation}\label{1}
\ddot{\Phi}+\Phi-|\Phi|\Phi+B{|\Phi|}^2\Phi+2s \Phi^{\ast}=0
\end{equation}
\begin{equation}\label{2}
j^0=\frac{1}{2i}({\Phi}^{\ast}\dot{\Phi}-\Phi{\dot{\Phi}}^{*})
\end{equation}
\begin{equation}\label{3}
\mathcal{E}=\frac12 |\dot{\Phi}|^2+\frac12 |\Phi|^2-\frac{1}{3}{|\Phi|}^3
+\frac{B}{4}{|\Phi|}^4+\frac s2 (\Phi^2+\Phi^{\ast 2})
\end{equation}
The absence of spatial derivative terms in the energy makes the
soliton more stable. We have found that on the boundary, the beat
frequency is proportional to the only free parameter of the
(rescaled) Lagrangian, namely $B$, taking care always to
keep the breaking parameter $s$ low enough to avoid the
energy/charge ratio to become greater than 1.
We have also checked that the same behaviour persists when we add a
$\Phi^6$ term in the $U(1)$ preserving potential.

\begin{figure}[t]
\centering
\includegraphics[scale=0.6]{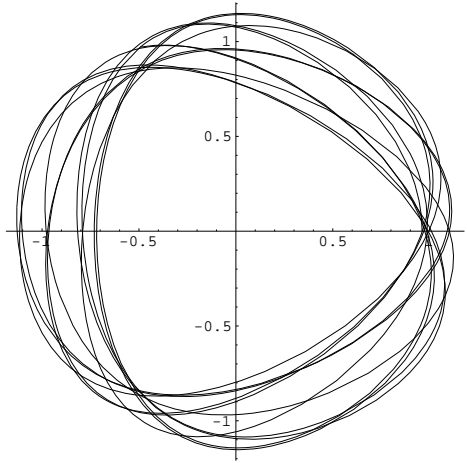}
\caption{Parametric plot (Real Part, Imaginary Part), in the case
where the breaking term is $\Phi^3+\Phi^{*3}$ and
for $x=0$, $s=0.05$ and for $t=100$}
\label{fig4}
\end{figure}

A last point to mention is the tremendously different situation we
are facing in the case we change the power of the symmetry
breaking term Eq.(\ref{breaking_t}) in the Lagrangian to $n=3$.
In Fig.(\ref{fig4}) we show the parametric plot for the same values
of the parameters, $B=0.34$, $\omega^2=0.51$ and for $s=0.05$.
Although the graph seems chaotic with respect to the corresponding
one in the previous case, we clearly see three ``fixed points''. A
larger $s$ value is also needed, with respect to the previous case
again, in order to see a significant disturbance of the soliton
profile. There appears again a $\Phi(x,t) $ which is  periodic in time, but
it does not show a beat-like shape and the period seems smaller
than before. We are planning to make a more general presentation
of these points in a forthcoming publication.

In conclusion, motivated by the $U(1)$ breaking symmetry terms that
appear in the soft-term Lagrangian, we investigated the simple case of
a complex scalar field, in 1+1 dimensions, possessing a
non-topological solition-type solution, with an extra
explicitly breaking term. We solve
numerically the equation of motion and found an oscillating with
time qball-like solution. The points of the profile oscillate
with a beat-like movement showing stability in time. As
the breaking parameter gets larger and the oscillating
energy/charge ratio gets greater than 1, instability appears
in the solution in the sense of non constant period and with the
destruction of the beat-like shape.

\vspace*{1cm}
We would like to thank  P. Dimopoulos, K. Farakos, A.
Kehagias and G. Koutsoumbas, G. Tiktopoulos for helpful discussions.


\end{document}